\documentclass[journal]{IEEEtran}

\usepackage{xcolor,soul,framed} 

\colorlet{shadecolor}{yellow}
\usepackage[pdftex]{graphicx}
\graphicspath{{../pdf/}{../jpeg/}}
\DeclareGraphicsExtensions{.pdf,.jpeg,.png}

\usepackage[cmex10]{amsmath}
\usepackage{array}
\usepackage{mdwmath}
\usepackage{mdwtab}
\usepackage{eqparbox}
\usepackage{url}
\usepackage{tikz}
\usepackage{caption}
\usepackage{booktabs}
\usepackage{amsthm}
\usepackage{multirow}
\usepackage{amsmath,amssymb,amsfonts}
\usepackage{float}
\usepackage{makecell}

\usepackage{adjustbox} 
\begin{document}
\bstctlcite{IEEEexample:BSTcontrol}
    \title{Physics-Informed Neural Operator for Electromagnetic Inverse Scattering Problems}
  \author{Q. C. Dong,~\IEEEmembership{Graduate Student Member,~IEEE,}  Zi-Xuan Su, Qing Huo Liu,~\IEEEmembership{Fellow,~IEEE,} Wen Chen,~\IEEEmembership{Senior Member,~IEEE,} Zhizhang (David) Chen,~\IEEEmembership{Fellow,~IEEE}

  \thanks{Manuscript received xxxx; revised xxxx. This article is a preprint going to be submitted to IEEE journals. The authors solely developed all results and methods. And all rights reserved. This manuscript has not yet been peer-reviewed. This paper is an expanded version from the 2026 IEEE International MTT Symposia (IMS): RF Technology \& Techniques (RFTT) and RF Systems \& Applications (RFSA) Symposia, Boston, Massachusetts, USA, 7-12 June, 2026. (Corresponding authors: Zhizhang (David) Chen, Wen Chen).
  
Q. C. Dong is with the Department of Electrical and Electronic Engineering at the Hong Kong Polytechnic University, Hong Kong, China, and the Eastern Institute of Technology, Ningbo, China  (e-mail: qc.dong@connect.polyu.hk). 

Zhizhang (David) Chen and Zi-Xuan Su are with the Department of Electrical and Computer Engineering, Dalhousie University, Halifax, Nova Scotia, Canada.

Qing Huo Liu is with the Eastern Institute of Technology, Ningbo, China.

Wen Chen is with the Department of Electrical and Electronic Engineering, the Hong Kong Polytechnic University, Hong Kong, China. } }

\markboth{xxxx, VOL.xxx, NO.xxx, xxxx
}{xxx \MakeLowercase{\textit{et al.}}: xxx}


\markboth{xxxxxxxx,~Vol.~xx, No.~x, xxxx~xxxx}%
{Shell \MakeLowercase{\textit{et al.}}: A Sample Article Using IEEEtran.cls for IEEE Journals}
\maketitle


\begin{abstract}
This paper proposes a physics-informed neural operator (PINO) framework for solving inverse scattering problems, enabling rapid and accurate reconstructions under diverse measurement conditions. In the proposed approach, the dielectric property is represented as a learnable tensor, while a neural operator is employed to predict the induced current distribution. A hybrid loss function, consisting of the state loss, data loss and total-variation (TV) regularization, is constructed to establish a fully differentiable formulation for a joint optimization of network parameters and dielectric property. To demonstrate the framework's generality and flexibility, PINO is implemented using three representative neural operators, i.e., the Fourier Neural Operator (FNO), the enhanced Fourier Neural Operator (U-FNO) and the Factorized Fourier Neural Operator (F-FNO). Compared with conventional approaches, the proposed framework offers a simpler formulation and universal modeling capability, making it readily applicable to various measurement scenarios, including multi-frequency and phaseless inversion. Numerical simulations demonstrate that the proposed PINO achieves high accuracy and robust reconstruction across samples with and without phase information, under single-frequency and multi-frequency settings in the presence of noise. The results demonstrate that PINO consistently outperforms conventional contrast-source inversion (CSI) methods and provides an efficient, unified solution to complex electromagnetic inverse-scattering problems.
\end{abstract}

%
%

\section{Introduction}
\label{sec1:intro}
The electromagnetic inverse scattering problems aim to reconstruct the spatial profiles of dielectric properties and the structural characteristics of targets from limited field measurements \cite{chen2022unified,chen2018computationalEIS}. Electromagnetic inverse scattering problems arise in a wide range of applications, including source reconstruction \cite{Time-reversal1,Time-reversal2,Time-reversal3}, medical imaging \cite{GSR}, and geophysical exploration \cite{SongChaofullwaveforminversion,SongChaoinversion}. However, there is the nonlinear coupling between the scattered field and the unknown dielectric contrast, as well as the severe ill-posedness caused by sparse measurement data and noise contamination \cite{ZhouZiheng,CCSI}. Consequently, conventional inversion methods often suffer from long solution time, high computational cost, limited reconstruction accuracy, and instability.

Recently, machine learning techniques \cite{Dong_2026,XuKuiwenlearningassist,DuYutong} have been applied to electromagnetic inverse problems to accelerate the solution process. Existing work has primarily focused on training end-to-end neural network models to approximate the underlying mapping between inputs and outputs \cite{Xudongreview1,Xudongreview2}. For instance, Yao \emph{et al.}~\cite{HeMingTwo-Step} proposed a two-step deep learning approach for inverse scattering problems. A deep convolutional neural network (CNN) first reconstructs an initial permittivity image from the measured scattering data, and then another CNN refines this image to improve the reconstruction performance \cite{Huhaojie1,HuHaoJie2}.

To further improve reconstruction accuracy and computational efficiency, physical priors have been incorporated into learning-based frameworks \cite{ShanTaoNeuralBorn,GuoRuiPENN,SOM-Net,GuoXingyuePICSI}. In particular, the backpropagation (BP) technique \cite{BP}, a fast imaging technique, has been employed to generate an initial coarse reconstruction result from the measured field data. The coarse reconstruction can be further refined with restoration networks, such as U-Net \cite{BPS,U-net,DeepNIS} or generative adversarial networks (GANs) \cite{PGAN}. Nevertheless, purely data-driven models often suffer from limited generalization, which leads to degraded reconstruction performance on the testing data deviated from the training distribution.

In this paper, we propose a physics-informed neural operator (PINO) framework for electromagnetic inverse scattering problems. The proposed framework extends the classical contrast source inversion (CSI) method \cite{CSI,CSI+TV} by integrating neural operators \cite{Neural-Operator} and concepts inspired by Neural Radiance Fields (NeRF) \cite{NeRF}. Specifically, a neural operator is employed to model the nonlinear mapping from spatial coordinates to induced current distributions over the computational domain, while the unknown dielectric property is represented as a learnable tensor. A hybrid loss function, consisting of the data loss, state loss, and total variation (TV) regularization, is constructed to jointly optimize the parameters of the neural operator and the dielectric properties within a fully differentiable framework. The effectiveness of the proposed method is validated through extensive numerical experiments under different noise levels and frequency settings. Comparative studies with traditional CSI and other machine-learning-based methods demonstrate that the proposed framework achieves improved reconstruction accuracy and robustness against noise contamination. Furthermore, the influence of different neural operator architectures used, including the Fourier Neural Operator (FNO) \cite{FNO}, enhanced-Fourier Neural Operator U-FNO \cite{U-FNO}, and the Factorized Fourier Neural Operator (F-FNO) \cite{F-FNO}, on inversion accuracy and computational efficiency is systematically investigated.

The remainder of this paper is organized as follows. Section \ref{sec2:Problem} briefly reviews the electromagnetic inverse scattering problem. Section \ref{sec3:Proposed} presents the proposed PINO framework. Section \ref{sec4:results} reports numerical results under different conditions. Conclusions are drawn in Section \ref{sec5:Conclusion}.

\begin{figure}
    \centering
    \includegraphics[trim=9cm 3cm 8cm 1cm,
    clip,width=0.45\textwidth]
    {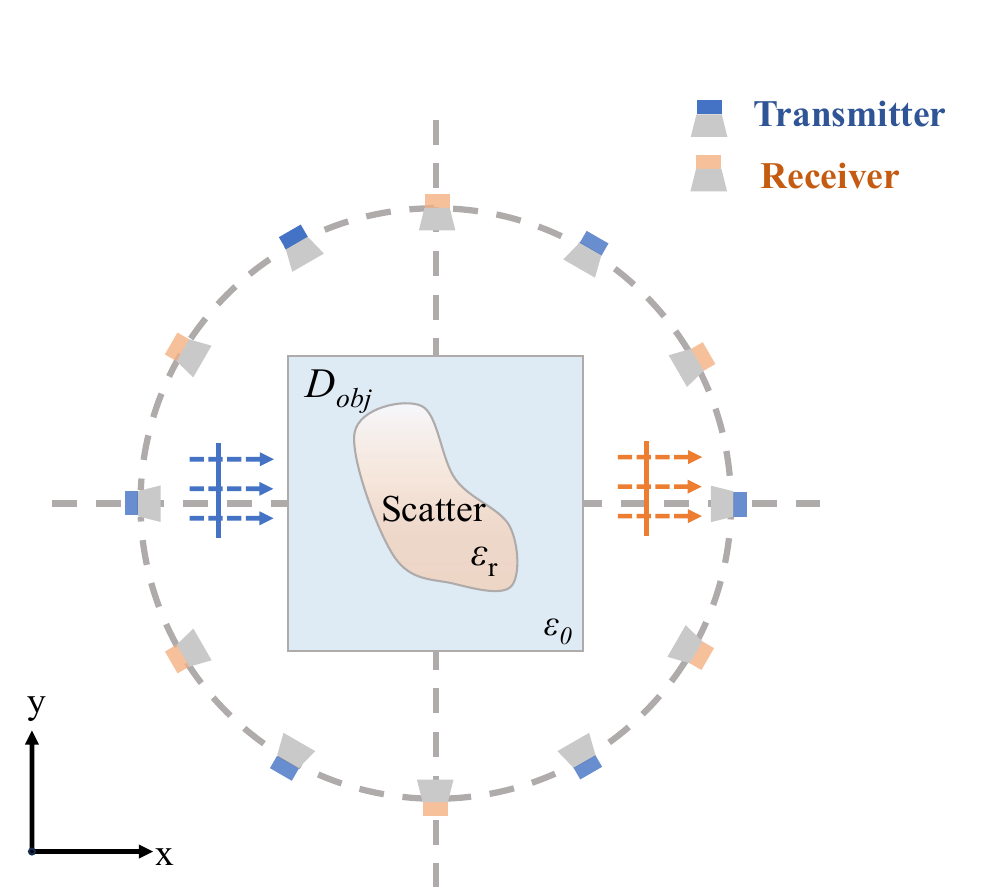}
    \centering
    \caption{A modeling setup: a two-dimensional TMz scattering scenario with an objective domain $D_{obj}$ containing a scatterer.}
    \label{fig:setup}
\end{figure}

\section{Problem statement}
\label{sec2:Problem}

\begin{figure*}[t]
    \centering
    \includegraphics[trim=0.2cm 2.7cm 0.2cm 1.2cm,
    clip,width=1.0\textwidth]{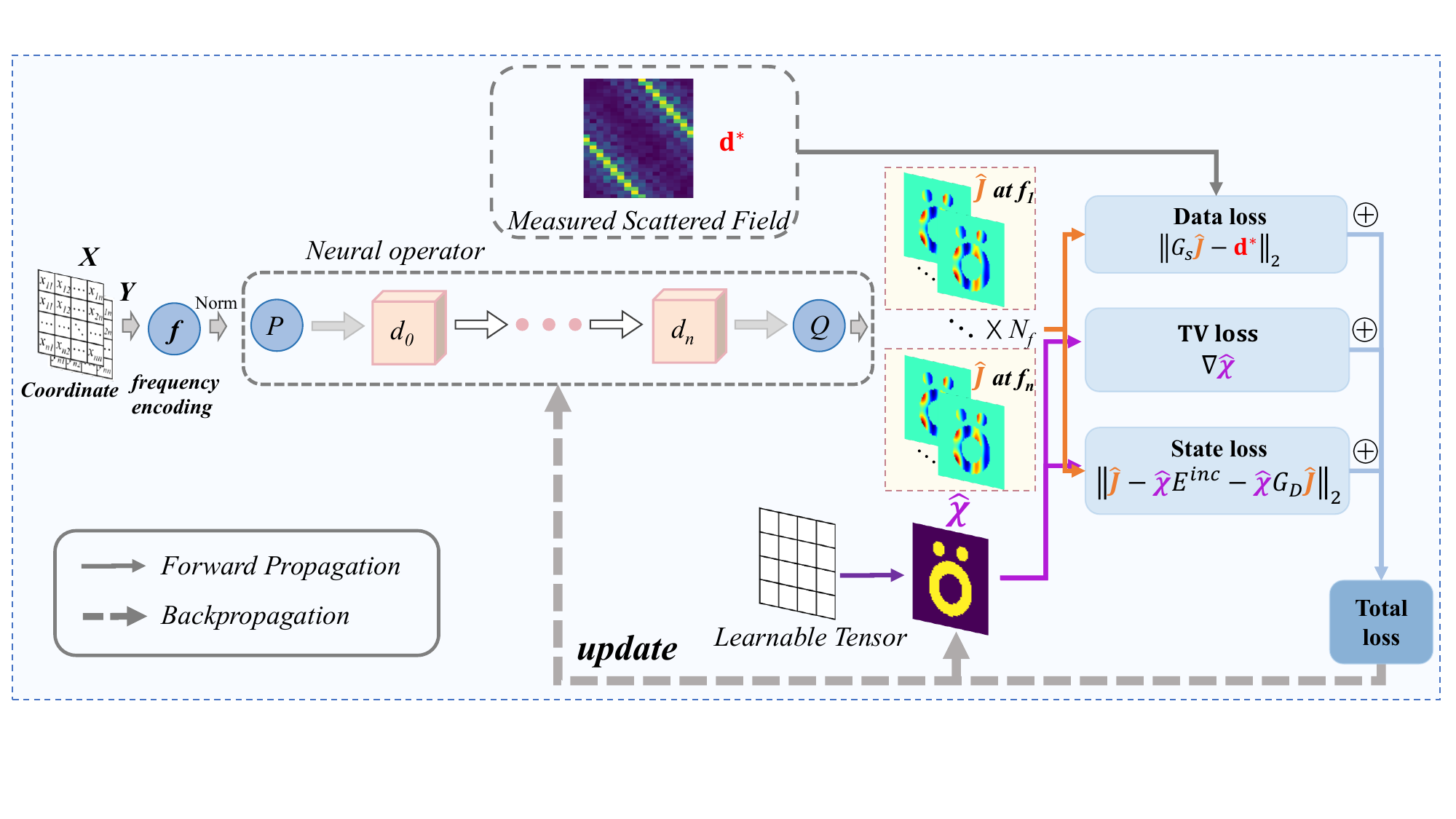}
    \caption{Framework of the proposed PINO for solving electromagnetic inverse scattering problem. The input consists of the normalized coordinates $X$ and $Y$, and the output of neural operator is the predicted induced current $\hat{J}$. The predicted contrast $\hat{\chi}$ is defined as one learnable tensor. $f_1$ and $f_n$ represent different frequencies.}
    \label{fig:framework}
\end{figure*}

As illustrated in Fig. \ref{fig:setup}, we consider a two-dimensional time-harmonic TM$z$ electromagnetic scattering problem. A scatterer with relative permittivity $\varepsilon_r$ is located within the object domain $D_{obj}$, which is embedded in a homogeneous background medium characterized by permittivity $\varepsilon_b$ and permeability $\mu_b$. The dielectric contrast function is defined as
$\chi = \varepsilon_r -\varepsilon_{r,b},$
where $\varepsilon_{r,b} = \varepsilon_b / \varepsilon_0$ denotes the relative permittivity of the background medium.

For a homogeneous background, the forward scattering problem can be formulated using the Lippmann–Schwinger integral equation \cite{Mom} described by

\begin{equation}
E(\mathbf{r}) = E^{inc}(\mathbf{r})
+ k_b^2\int_{D_{obj}} G(\mathbf{r},\mathbf{r'})\,\chi(\mathbf{r'})\,E(\mathbf{r'})\; d\mathbf{r'},
\label{eq:state}
\end{equation}
where $E^{\mathrm{inc}}$ denotes the incident field, $G(\mathbf{r},\mathbf{r'})$ is the two-dimensional Green's function, and $k_b$ denotes background wavenumber $\omega\sqrt{\varepsilon_b\mu_b}$. Here, $\mathbf{r}$
denotes the observation point and $\mathbf{r'}$ denotes the source position.

To formulate the inverse problem, the induced contrast current density is introduced as

\begin{equation}
J(\mathbf{r'})=\chi(\mathbf{r'})E(\mathbf{r'}),
\end{equation}
which allows the scattered field measured at the receivers to be written as
\begin{equation}
E^{sca}(\mathbf{r})
= k_b^2\int_{D_{obj}} G(\mathbf{r},\mathbf{r'})\, J(\mathbf{r'})\, d\mathbf{r'}.
\label{eq:scatter}
\end{equation}

By discretizing the object domain $D_{obj}$ \cite{chenMom}, the integral equations can be expressed in matrix forms as

\begin{equation}
    E^{tot} = E^{inc} + G_D \chi E^{tot},
    \label{eq:total_field}
\end{equation}

\begin{equation}
    E^{sca} = G_S \chi E^{tot},
    \label{eq:scattered_field}
\end{equation}
where ${G}_D$ and ${G}_S$ denote the discretized Green’s operators associated with the object domain and the observation domain, respectively.

The objective of the electromagnetic inverse scattering problem is to reconstruct the contrast function $\chi$ from the measured scattered-field data $\mathbf{d}^*$. This task is commonly formulated as the minimization of a loss function with the form described by
\begin{equation}
    L(\chi) = \| \mathbf{d}^* - \hat{E}^{sca} \|^2,
    \label{eq:cost_function}
\end{equation}
where $\hat{E}^{sca}$ denotes the scattered field simulated using the predicted contrast $\hat{\chi}$ through the forward scattering model described in~\eqref{eq:scattered_field}.  Owing to the nonlinear dependence of the scattered field on the contrast function $\chi$ and the sparsity of the measurement data, the resulting inverse problem is nonlinear and ill-posed \cite{chen2018computationalEIS}, which poses a significant challenge for achieving stable and accurate reconstructions.

\section{The Proposed Framework}
\label{sec3:Proposed}
\subsection{Neural Operator}

Neural operators~\cite{Dong_2026,FNO,PNO} have recently emerged as a powerful learning paradigm for modeling the mapping between infinite-dimensional function spaces especially for partial differential equations (PDEs). Unlike conventional neural networks that operate on discretized inputs and outputs, neural operators aim to learn the solution operators of PDEs in a resolution-invariant manner.

In this work, neural operator architectures are combined with physical constraints to develop the PINO framework for solving the inverse problems. In practice, the discretized coordinate matrices $X$ and $Y$ of the grid over $D_{obj}$, as illustrated in Fig.~\ref{fig:framework}, are used as inputs of the proposed framework.

To enhance the representation capability of spatial coordinates, frequency encoding is first applied to the coordinates. The encoding function $f$ follows the formulation adopted in \cite{NeRF,luo2024imaging,deepcsi} and is defined as
\begin{equation}
    f(X, Y) = 
    \begin{bmatrix}
        \cos(2X), \sin(2X), \ldots, \cos(2^M X), \sin(2^M X) \\
        \cos(2Y), \sin(2Y), \ldots, \cos(2^M Y), \sin(2^M Y)
    \end{bmatrix}^T.
\end{equation}

In this work, $M$ is set to 10. The neural operator first applies a linear lifting operator $P$ to project the encoded input $f(X,Y)$ into a higher-dimensional feature space as described by

\begin{equation}
    {d}_0(\mathbf{x}) = P( f(X, Y)),
\end{equation}
where $\mathbf{x} \in D_{obj}$ denotes the spatial coordinates in the physical domain, and $ P $ is typically parameterized by a fully-connected neural network (FCNN).

Within the neural operator framework, $ {d}_0 $ serves as the initial feature field, which is successively transformed through a sequence of iterative operator layers, i.e., ${d}_0 \rightarrow {d}_1 \rightarrow \cdots \rightarrow {d}_n$. Here, ${d}_i(\mathbf{x})$ represents the hidden feature field at the $i$-th layer, where intermediate representations are progressively updated. The specific operations within each layer vary across different neural operator models \cite{FNO,MGNO,WNO}. Here, we illustrate the procedure using the FNO as a representative example. Specifically, the core iterative update in FNO transforms the representation state from ${d}_i$ to ${d}_{i+1}$ according to
\begin{equation}
    {d}_{i+1}(\mathbf{x}) = \sigma\left( W {d}_i(\mathbf{x}) + (\mathcal{K}(\phi){d}_i)(\mathbf{x}) \right), i \geq 0.
    \label{eq:fno_layer}
\end{equation}
where $ \sigma(\cdot)$ denotes a nonlinear activation function and $W$ is a learnable matrix that performs a linear transformation. The operator $\mathcal{K}$ is the kernel integral operator defined as

\begin{equation}
    (\mathcal{K}(\phi){d}_i)(\mathbf{x}) = \mathcal{F}^{-1} \left[ \mathcal{F}[\kappa_\phi](\mathbf{x}) \cdot \mathcal{F}[{d}_i](\mathbf{x}) \right],
    \label{eq:fno_fourier}
\end{equation}
where $ \mathcal{F} $ and $ \mathcal{F}^{-1} $ denote the forward and inverse Fourier transforms, respectively. The function $\kappa_\phi$ represents a neural network parameterized by $\phi$. Instead of explicitly computing $\mathcal{F}[\kappa_\phi]$, a low-pass learnable complex-valued tensor $\mathbf{R}$ is introduced to directly parameterize the Fourier kernel.

Each update step in~\eqref{eq:fno_layer} corresponds to one FNO layer. After several iterations, the resulting feature field ${d}_n$ is fed into the FCNN, and another linear operator $Q$ projects the output to the target dimension of the predicted induced current distributions $\hat{J}$:
\begin{equation}
    \hat{J} = Q({d}_n),
\end{equation}
where $\hat{J}$ represents the predicted induced current distributions.

\subsection{Inversion using With-phase Data}
For the inverse scattering problem, the proposed framework employs one neural operator to represent the mapping described by

\begin{equation} \mathcal{F}^{inv}_{\theta}: f(X,Y) \longmapsto\ \hat{J}, \end{equation} where $\hat{J}$ is the predicted induced current corresponding to different frequencies shown in Fig. \ref{fig:framework}. In addition, a learnable tensor is introduced to represent the predicted dielectric contrast $\hat{\chi}$.

The state loss $L_{state}$ is defined as

\begin{equation}
L_{state}= \frac{1}{N_f} \sum_{i=1}^{N_f}\frac{\big\|\hat{J}-\hat{\chi} E^{inc}-\hat{\chi} G_D \hat{J}\big\|_2}{\big\|\hat{\chi} E^{inc}\big\|_2},
\end{equation} 
which is derived from (\ref{eq:state}). Here, $N_f$ denotes the number of frequencies and $i=1,\dots,N_f$ indexes the frequency samples. The state loss is normalized by 
${\big\|\hat{\chi} E^{inc}\big\|_2}$. When this quantity approaches zero (e.g., when $\hat{\chi}$ is close to zero), the normalization can be stabilized using a fixed scale based on the incident field. 

The data loss $L_{data}$ is formulated based on (\ref{eq:scatter}) as described by
\begin{equation}
L_{data}=\frac{1}{N_f} \sum_{i=1}^{N_f}\frac{\big\|G_S\hat{J}-\mathbf{d}^{*}\big\|_2}{\big\|\mathbf{d}^{*}\big\|_2}. \label{eq:dataloss}
\end{equation}

The optimization is performed by minimizing the total loss $L_{total}$ following \cite{luo2024imaging,deepcsi}:
\begin{equation}
\min_{\theta}L_{total}=\lambda_1 L_{state}+\lambda_2L_{data}+\lambda_R\|\nabla \hat{\chi}\|_{1/2},
\label{total loss}
\end{equation}

Here $\|\nabla \hat{\chi}\|_{1/2}$ denotes the fractional-order total variation (TV) loss in Fig. \ref{fig:framework}, and $\nabla$ is the discrete gradient operator used to enforce the smoothness of the reconstructed contrast. The hyperparameters $\lambda_1$, $\lambda_2$, and $\lambda_R$ balance the contributions of different loss terms. The variable $\theta$ denotes the trainable parameters of the neural operator $\mathcal{F}^{inv}_{\theta}$ together with the learnable tensor $\hat{\chi}$.

\subsection{Inversion using Phaseless Data}

In practice, accurate phase information is difficult to obtain, particularly at high frequencies. As a result, conventional inverse scattering methods often require a reformulation when only intensity data are available.

In contrast, the proposed PINO framework is fully differentiable, which enables the implementation of phaseless data (PD) inverse scattering using a modified loss function. For phaseless inversion, the contrast function is constrained to be non-negative to mitigate the strong nonlinearity introduced by the loss of phase information. This constraint is achieved by using a differentiable re-parameterization described by
\begin{equation}
\hat{\chi}_{pd}=\operatorname{softplus}(\hat{\chi}),
\end{equation}
where $\mathrm{softplus}(\cdot)$ is a smooth activation function to guarantee $\hat{\chi}_{pd}>0$.

Under phaseless measurement conditions, only the magnitude of the total electric field is available at the receivers. Accordingly, the state and data loss functions are reformulated as
\begin{equation}
L^{pd}_{state}=\frac{1}{N_f} \sum_{i=1}^{N_f}\frac{\big\|\hat{J}-\hat{\chi}_{pd} E^{inc}-\hat{\chi}_{pd} G_D \hat{J}\big\|_2}{\big\|\hat{\chi}_{pd} E^{inc}\big\|_2},
\end{equation}
\begin{equation}
L^{pd}_{data}=\frac{1}{N_f} \sum_{i=1}^{N_f}\frac{\big\||E_r^{inc} + G_S \hat{J}|-\mathbf{d}^{*}\big\|_2}{\big\|\mathbf{d}^{*}\big\|_2},
\end{equation}
where $E_r^{inc}$ denotes the incident field directly radiated from the transmitter to the receiver.

The optimization process of phaseless data inversion is carried out by minimizing the following loss function:
\begin{equation}
\begin{aligned}
\min_{\theta}L^{pd}_{total} &= \lambda_1L_{state}^{pd} +\lambda_2L_{data}^{pd} + \lambda_R\|\nabla \hat{\chi}_{pd}\|_{1/2}
.
\end{aligned}
\label{eq:loss_pd}
\end{equation}

\section{Numerical results}
\label{sec4:results}

In the numerical experiments, the object domain $D_{obj}$ is defined as a square region of size $0.3\times 0.3$ $\text{m}^2$. A total of $N_i = 16$ line sources are employed for illumination and $N_s = 32$ receiving antennas are used to collect the scattered fields. Both transmitters and receivers are uniformly distributed at a distance of $2$ $\text{m}$ from the center of the object domain $D_{obj}$. The domain $D_{\mathrm{obj}}$ is discretized into $128\times128$ grids
for the forward modeling process using the Method of Moments (MoM) \cite{Mom,chenMom2}. To avoid the inverse crime \cite{inversecrime}, all inversion procedures are performed on a coarser $64 \times 64$ grid. 

To simulate practical measurement uncertainties, additive noise is introduced to the scattered field $E^{sca}$. Since the scattered field is complex-valued, independent additive Gaussian white noise is applied to the real and imaginary parts, respectively. The noisy scattered field is generated by
\begin{equation}
\operatorname{Re}(E_{noisy}^{sca}) = \operatorname{Re}(E^{sca}) + a \, |\operatorname{Re}(E^{sca})| \, \mathcal{N}(0,1),
\end{equation}
\begin{equation}
\operatorname{Im}(E_{noisy}^{sca}) = \operatorname{Im}(E^{sca}) + a \, |\operatorname{Im}(E^{sca})| \, \mathcal{N}(0,1),
\end{equation}
where $\mathcal{N}(0,1)$ denotes a zero-mean, unit-variance Gaussian random variable, and $a$ represents a noise level expressed as a specific percentage.

In numerical experiments, the reconstruction quality is evaluated by comparing the recovered relative permittivity $\hat{\epsilon}_r$ with the ground truth $\epsilon_r$ using the
structural similarity index measure (SSIM) and normalized root mean square error (RMSE) defined in Appendix.
All numerical experiments are conducted on a cloud instance with 18 vCPUs (backed by an AMD EPYC 9754 processor) and a single NVIDIA RTX 4090D GPU.

\subsection{Single Frequency With-phase Data Inversion From Scratch}

To evaluate the effectiveness of the proposed framework, we assess the reconstruction performance on three representative targets, namely the Austria-shaped target, the cylindrical target, and the digit-shaped target selected from MNIST dataset \cite{MNIST}. The inversion is performed using the FNO-based implementation of the proposed PINO framework at a single frequency of $3$ $\mathrm{GHz}$. The FNO employs 20 Fourier modes with a hidden width of 64. Training is conducted using the Adam optimizer \cite{Adam} for 2{,}000 epochs, with an initial learning rate of $5\times 10^{-3}$ that decays by a factor of 0.5 every 400 epochs.

\begin{table}[b]
\centering
\caption{Inversion results at 3 GHz inversion with 5\% noise levels with FNO.}
\label{tab:exp1_singlefreq_FNO}
\begin{tabular}{ccc}
\hline
Case & SSIM & RMSE\\
\hline
(a) Austria & 0.9163 & 0.0473\\
(b) Cylinders & 0.9533 & 0.0263\\
(c) MNIST & 0.9133 & 0.0618\\
\hline
\end{tabular}
\end{table}

\begin{figure*}
    \centering
    \includegraphics[trim=0cm 0cm 0cm 0cm,
    clip,width=1\textwidth]
    {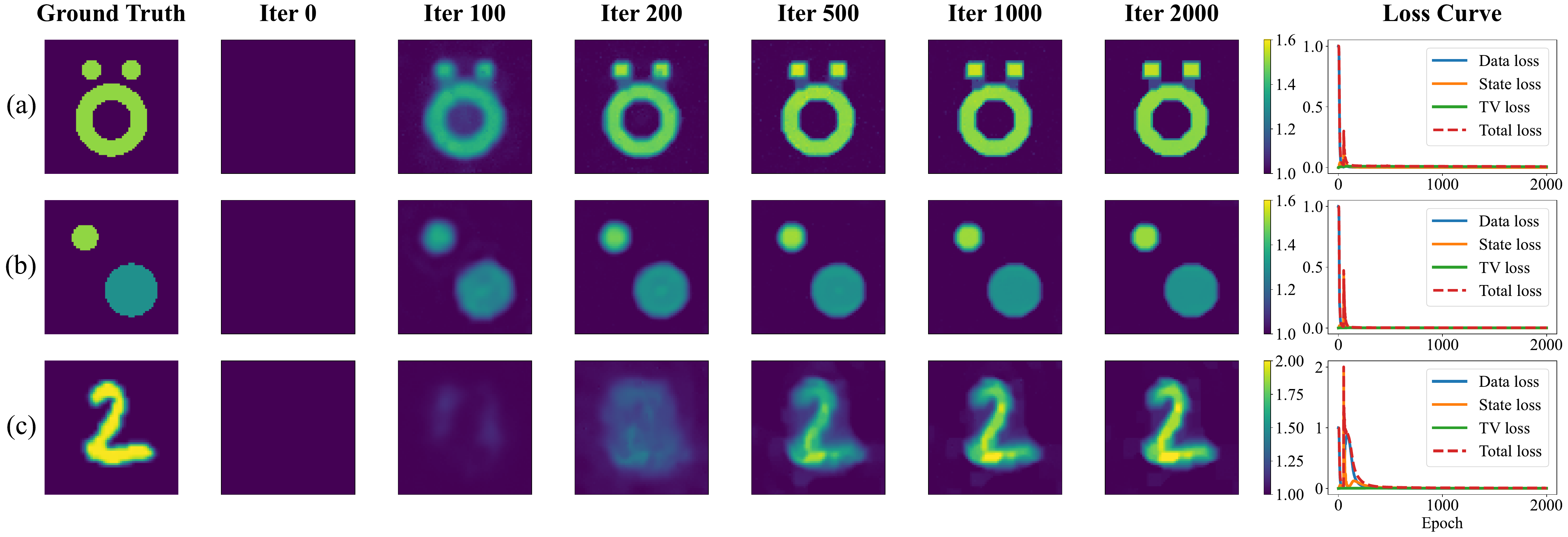}
    \centering
    \caption{(a)--(c) Ground truth and corresponding inversion results of the proposed framework with FNO at 3 GHz with 5\% noise. ‘Iter $n$’ represents the inversion results after the $n$-th iteration. The rightmost column represents the loss curves during the iteration process.}
    \label{fig:SF_total_result}
\end{figure*}

\begin{table*}
\centering
\caption{Inversion results obtained by using different methods at 3 GHz.}

\begin{tabular}{cccccccccccccccc}
\toprule
\multirow{2}{*}{\makecell{Noise\\level}} & \multicolumn{2}{c}{Proposed} & \multicolumn{2}{c}{CSI} & \multicolumn{2}{c}{CSI+TV} & \multicolumn{2}{c}{BP} & \multicolumn{2}{c}{BPS} & \multicolumn{2}{c}{Physics-Net} & \multicolumn{2}{c}{CS-Net} \\
\cmidrule(lr){2-3} \cmidrule(lr){4-5} \cmidrule(lr){6-7} \cmidrule(lr){8-9} \cmidrule(lr){10-11} \cmidrule(lr){12-13} \cmidrule(lr){14-15}
 & SSIM & RMSE & SSIM & RMSE & SSIM & RMSE & SSIM & RMSE & SSIM & RMSE & SSIM & RMSE & SSIM & RMSE \\
\midrule
10\% & \textbf{0.9166}
 & \textbf{0.0488} & 0.6180 & 0.1253 & 0.6194 & 0.1243 & 0.5630
 & 0.1483 & 0.8142 & 0.0806 & 0.8522
 & 0.0625 & 0.7913
 & 0.0651\\
30\% & \textbf{0.8719}
 & \textbf{0.0644} & 0.5819 & 0.1333
 & 0.5952 & 0.1263 & 0.5554
 & 0.1497 & 0.7455 &0.08712
 & 0.7957
 & 0.0723 & 0.4724
 & 0.1250\\
50\% & \textbf{0.7651}
 & \textbf{0.0859} & 0.5089
 & 0.1539 & 0.5444 & 0.1316 & 0.5398
 & 0.1507 & 0.6392
 & 0.1040 & 0.7063
 & 0.0861 & 0.4922
 & 0.0934\\
\bottomrule
\end{tabular}
\label{exp2_quant}
\end{table*}

Fig.~\ref{fig:SF_total_result} shows the ground truth and reconstructed results under $5\%$ Gaussian noise at different iterations of the inversion process. The corresponding evolution of data loss, state loss, and TV loss is also presented. Starting from a homogeneous initial guess, satisfactory reconstructions are obtained after approximately 500 iterations. As the inversion proceeds, the reconstructed images are progressively refined, and high-quality reconstructions are achieved after 2{,}000 iterations. Quantitative results in terms of SSIM and RMSE of the reconstructed results and ground truths are given in Table~\ref{tab:exp1_singlefreq_FNO}. The reconstructed results consistently achieve SSIM values above 0.9 and RMSE values below 0.1, indicating high-quality reconstruction.

This experiment is designed primarily to demonstrate the effectiveness of the proposed framework when initialized from scratch. In practical inversion scenarios, direct inversion methods, such as BP \cite{BP} and time-reversal \cite{Time-reversal2}, can be used to provide informative prior reconstructions. A better initial guess can reduce the number of iterations and accelerate inversion convergence. Furthermore, transfer learning techniques in machine learning can further improve the efficiency of the inversion process \cite{Dong_2026}.

\subsection{Comparison with the State-of-the-Art Methods}
\label{sec:exp2}

In this section, we evaluate the robustness of PINO with an FNO backbone against different noise levels under single-frequency with-phase data conditions.
All methods are evaluated using the data collected at a fixed frequency of 3 GHz with additive noise levels of 10\%, 30\%, and 50\%. 
To ensure a fair comparison, all competing approaches share the same forward model, measurement configuration, and identical noise realizations.

The proposed method is compared with several representative inverse scattering methods, including traditional approaches such as CSI \cite{CSI}, CSI+TV \cite{CSI+TV}, and BP \cite{BP}, as well as machine learning-based methods including BPS \cite{BPS}, Physics-Net \cite{Physics-net}, and CS-Net \cite{CS-net}. The BPS method employs U-Net to refine the coarse reconstruction obtained from BP, while Physics-Net incorporates a hybrid data-driven and physics-informed loss to improve BP-based inversion results. The CS-Net can be viewed as an enhanced variant of the subspace-based optimization method (SOM) \cite{SOM}. It employs a neural network to estimate the noise subspace components of the contrast source, thereby improving the initialization for subsequent SOM+TV optimization. All baseline methods are implemented via following their original works and available official releases.

\begin{figure*}
    \centering
    \includegraphics[trim=0cm 0cm 0cm 0cm,
    clip,width=0.96\textwidth]
    {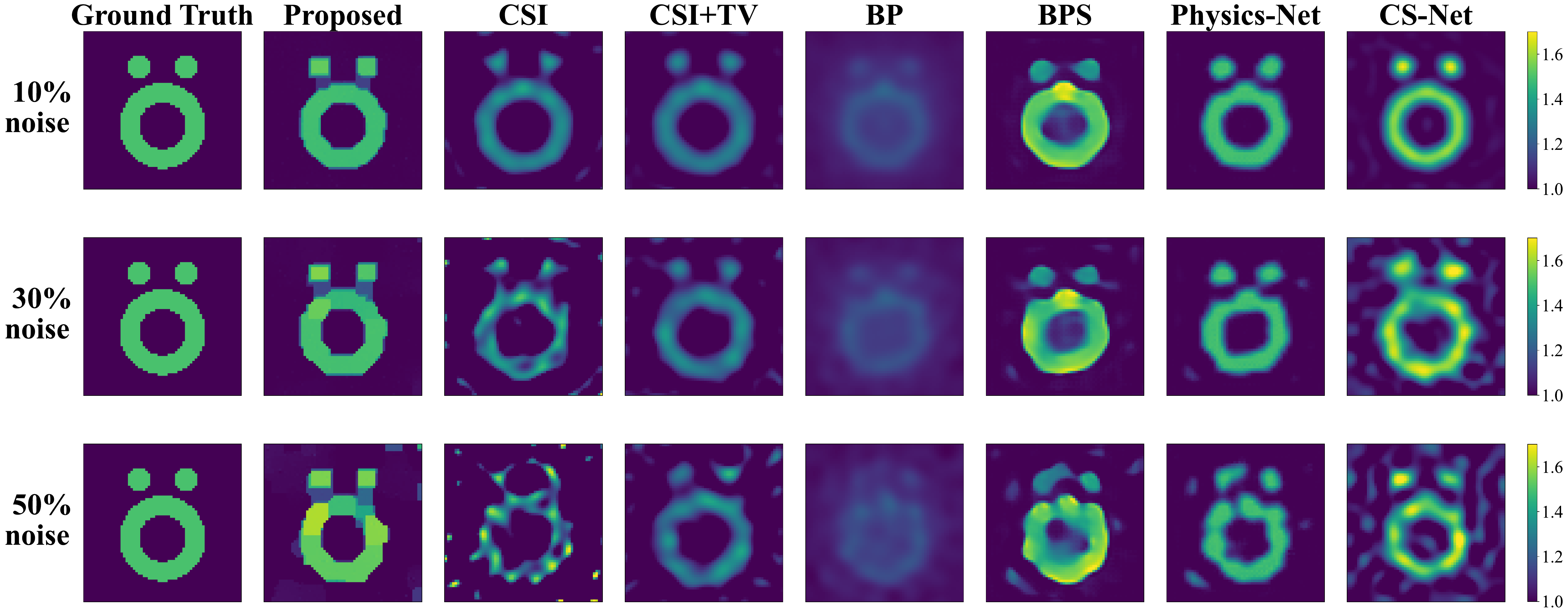}
    \caption{Comparison of inversion results under 3 GHz using different methods.}
    \label{fig:exp2_singlefreq_compare}
\end{figure*}

Comprehensive comparison results are given in Table~\ref{exp2_quant} and Fig.~\ref{fig:exp2_singlefreq_compare}, and FNO consistently achieves the highest SSIM and the lowest RMSE across all tested noise levels. Under 10\% noise, traditional methods, such as CSI, CSI+TV, and CS-Net, tend to produce reconstructions with blurred or ambiguous boundaries, while the fast BP method exhibits relatively large reconstruction errors. For learning-based baselines, including BPS and Physics-Net, performance degrades when applied to unseen noisy data, reflecting their limited generalization and strong dependence on initialization quality of the BP method. At the higher noise levels (50\% noise), competing methods suffer from pronounced artifacts and structural distortions, whereas FNO maintains stable reconstruction quality. These results demonstrate that the proposed framework achieves more accurate and stable reconstructions under severely ill-posed and noisy single-frequency measurement conditions.

\begin{table*}
\centering
\caption{With-phase data inversion results using different neural operators}
\label{tab:experiment3}
\small
\begin{tabular}{c c ccc ccc ccc}
\toprule

& & \multicolumn{3}{c}{FNO} 
& \multicolumn{3}{c}{U-FNO} 
& \multicolumn{3}{c}{F-FNO} \\
\cmidrule(lr){3-5}
\cmidrule(lr){6-8}
\cmidrule(lr){9-11}
Noise level & Frequency & SSIM $\uparrow$  & RMSE $\downarrow$ & Time$^{1}$ $\downarrow$
& SSIM $\uparrow$ & RMSE $\downarrow$ & Time$^{1}$ $\downarrow$
& SSIM $\uparrow$& RMSE $\downarrow$ & Time$^{1}$ $\downarrow$\\
\midrule
10\%&3 GHz&\textbf{0.91181}&\textbf{0.0486}&\textbf{11.55}&0.91178&0.0498&21.60&0.8945&0.0579&17.71\\
10\%&3,4,5 GHz&0.9279&0.0411&13.54&\textbf{0.9314}&\textbf{0.0404}&19.20&0.9190&0.0470&\textbf{13.21}\\
50\%&3 GHz&0.7446&0.0871&\textbf{12.34}&0.7488&0.0872&22.22&\textbf{0.8082}&\textbf{0.0814}&18.00\\
50\%&3,4,5 GHz&\textbf{0.9040}&\textbf{0.0582}&\textbf{11.67}&0.8923&0.0620&18.60&0.8526&0.0809&15.78\\
\bottomrule
\end{tabular}
\caption*{$^{1}$ unit: ms/iteration. $\uparrow$: higher is better. $\downarrow$: lower is better. Bold indicates the best result.}
\end{table*}

\begin{figure}
    \centering
    \includegraphics[trim=0.1cm 0cm 0cm 0cm,
    clip,width=0.48\textwidth]
    {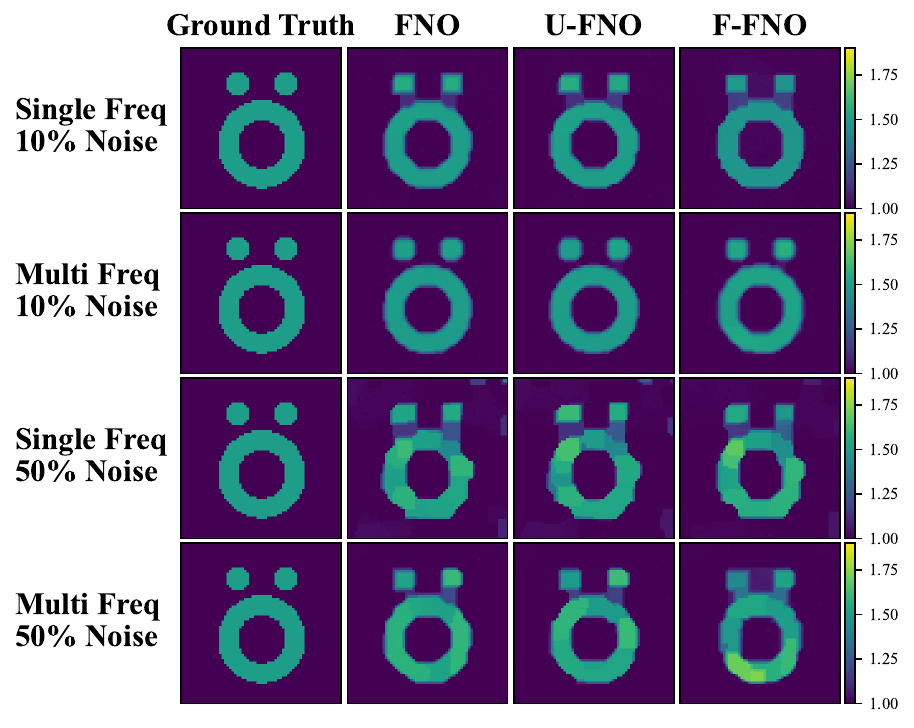}
     \caption{Comparison of with-phase data inversion results using different neural operators.}
    \label{fig:exp3_backbone}
\end{figure}

\subsection{With-phase Data Inversion using Different Neural Operators}
\label{sec:exp3}

The proposed framework, based on automatic differentiation, can be naturally extended to multi-frequency inversion scenarios. This extension only requires modifying the output of the neural operator to represent induced current distributions at different frequencies. In this section, we also investigate the impact of different neural operator backbones within the proposed PINO framework.

With the introduction of FNO, numerous variants have been proposed to enhance its performance, including U-FNO and F-FNO. U-FNO integrates an additional U-Net architecture into the Fourier layers to improve the expressive capability of the original FNO. F-FNO incorporates tensor factorization techniques to impose structural constraints on the weight tensors, thereby reducing model complexity and parameter count. In this study, FNO, U-FNO, and F-FNO are selected as representative backbones for a fair comparison, with inversion performed under different noise levels and frequency settings.

Experiments are conducted under both single-frequency ($3$ $\mathrm{GHz}$) and multi-frequency settings ($3,4,5$ $\mathrm{GHz}$), with noise levels of $10\%$ and $50\%$. All models share the same training protocol, loss formulation, and noise contamination. The inversion results are summarized in Table~\ref{tab:experiment3} and Fig.~\ref{fig:exp3_backbone}. The results indicate that no single neural operator backbone consistently dominates across all evaluation metrics and measurement conditions. Instead, different architectures exhibit complementary strengths under different noise levels and frequency settings. Specifically, FNO strikes a favorable balance between reconstruction accuracy and computational efficiency, achieving competitive SSIM and RMSE values while incurring the lowest per-iteration computational time in most cases. U-FNO attains the highest reconstruction accuracy under $10\%$ noise in multi-frequency settings, whereas F-FNO demonstrates improved robustness under $50\%$ noise in single-frequency inversion.

\begin{table}
\centering
\caption{Parameters of different models}
\label{tab:Parameter of different models}
\small
\begin{tabular}{c c c c}
\toprule

&\multicolumn{1}{c}{FNO} 
&\multicolumn{1}{c}{U-FNO} 
&\multicolumn{1}{c}{F-FNO}\\
\cmidrule(lr){2-2}
\cmidrule(lr){3-3}
\cmidrule(lr){4-4}
 Frequency &Parameters$^{2}$ $\downarrow$
&Parameters$^{2}$$\downarrow$
& Parameters$^{2}$ $\downarrow$ \\
\midrule
3 GHz   &13.14 &20.65  &\textbf{0.23} \\
3,4,5 GHz  &13.15 &20.66  &\textbf{0.24}\\
\bottomrule
\end{tabular}
\caption*{$^{2}$ unit : Mb.}
\end{table}

\begin{table*}[t]
\centering
\caption{Phaseless-data inversion results using different neural operators}
\label{tab:comparison}
\small
\begin{tabular}{c c ccc ccc ccc}
\toprule

& & \multicolumn{3}{c}{FNO} 
& \multicolumn{3}{c}{U-FNO} 
& \multicolumn{3}{c}{F-FNO} \\
\cmidrule(lr){3-5}
\cmidrule(lr){6-8}
\cmidrule(lr){9-11}
Noise level & Frequency & SSIM $\uparrow$  & RMSE $\downarrow$& Time$^{1}$ $\downarrow$
& SSIM $\uparrow$ & RMSE $\downarrow$& Time$^{1}$ $\downarrow$
& SSIM $\uparrow$& RMSE $\downarrow$ & Time$^{1}$ $\downarrow$\\
\midrule
10\%&3 GHz &0.8639 &\textbf{0.0566} &\textbf{11.94}& \textbf{0.8695}& 0.0604&21.96&0.8493 &0.0581 &17.22\\
10\%&3,4,5 GHz &\textbf{0.9070} & \textbf{0.0505}&\textbf{13.16} &0.9060& 0.0543&19.50&0.8922 &0.0548 &15.27\\
50\%&3 GHz &0.7214 &\textbf{0.0871}&\textbf{11.66} &\textbf{0.7346} &0.0933&21.23&0.7003 &0.0993 &17.55\\
50\%&3,4,5 GHz &\textbf{0.8289} &\textbf{0.0741}&\textbf{13.56} & 0.8218&0.0760 &17.23&0.7981 & 0.0830&14.34\\
\bottomrule
\end{tabular}
\caption*{$^{1}$ unit: ms/iteration.}
\end{table*}

The parameter statistics of different models are presented in Table~\ref{tab:Parameter of different models}. Due to factorized Fourier representation, F-FNO contains the fewest parameters and requires the least memory. However, this design also introduces additional tunable hyperparameters, making it challenging in practice to achieve optimal performance through empirical hyperparameter tuning alone. Since U-FNO employs additional convolutional layers, its parameter count exceeds that of the original FNO. Although this design can enhance reconstruction accuracy under specific conditions, it also results in a significantly longer per-iteration runtime compared with the other two backbones, as shown in Table~\ref{tab:experiment3}. Importantly, all backbones yield stable and reliable reconstructions within the proposed PINO framework, which confirms that the proposed framework is largely backbone-agnostic. Moreover, for all noise levels, multi-frequency inversion consistently outperforms single-frequency inversion across all three backbones.

\subsection{Phaseless Data Inversion using Different Neural Operators}
\label{sec:exp4}

In practical measurement processes, accurately measuring phase data is often expensive. Extending traditional inverse scattering methods to phaseless inversion typically requires re-deriving and modifying the inversion formulation. In contrast, the proposed framework supports single-frequency and multi-frequency phaseless inversion through a direct modification of the data loss function. We further evaluate the proposed PINO framework under the challenging phaseless settings, where only intensity information is available.

\begin{figure}[t]
    \centering
    \includegraphics[trim=0.1cm 0cm 0cm 0cm,
    clip,width=0.48\textwidth]
    {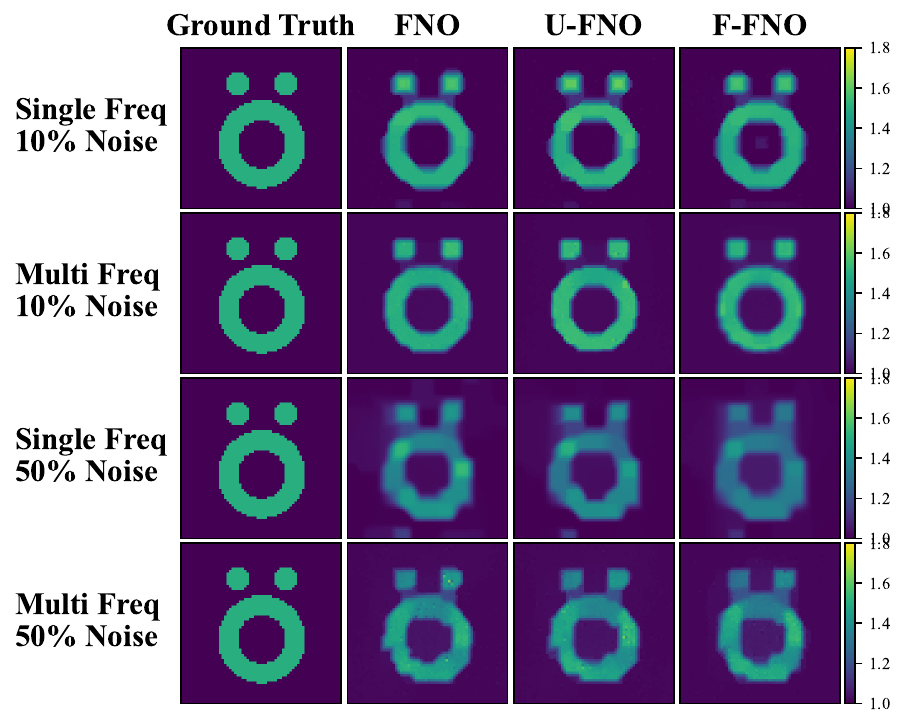}
    \caption{Comparison of phaseless data inversion results using different neural operators.}
    \label{fig:exp4_phaseless}
\end{figure}

We consider both single-frequency (3 GHz) and multi-frequency (3, 4, 5 GHz) phaseless inversion scenarios with noise levels of $10\%$ and $50\%$.
All three backbones (FNO, U-FNO, F-FNO) are evaluated under identical conditions. Compared with with-phase data inversion, the neural operator configurations remain unchanged, and the corresponding hyperparameters follow Table~\ref{tab:Parameter of different models}. As shown in Fig.~\ref{fig:exp4_phaseless}, the proposed framework achieves relatively accurate reconstructions across different settings. Quantitative evaluation results of phaseless inversion are summarized in Table~\ref{tab:comparison}. Despite the absence of phase information, all three backbones integrated into the proposed PINO framework produce stable and meaningful reconstructions under both $10\%$ and $50\%$ noise levels. Moreover, consistent with the with-phase data case, multi-frequency inversion substantially improves reconstruction accuracy compared with single-frequency inversion across all backbone architectures, which highlights the role of frequency diversity in mitigating the ill-posedness of phaseless inverse scattering.

Among the three backbones, FNO generally provides a favorable balance between reconstruction accuracy and computational efficiency, yielding competitive SSIM and RMSE values with the lowest per-iteration runtime in most scenarios. U-FNO occasionally achieves slightly higher reconstruction accuracy at the expense of increased computational cost, whereas F-FNO achieves the smallest parameter count but exhibits reduced accuracy under phaseless conditions. Importantly, despite these architectural differences, all three backbones can be effectively incorporated into the proposed PINO framework, which confirms the robustness and effectiveness of the proposed approach under highly challenging phaseless measurement conditions.

\section{Conclusion}
\label{sec5:Conclusion}

This paper has presented a unified PINO framework for electromagnetic inverse scattering problems. In the proposed approach, a neural operator is employed to predict induced current distributions, while the dielectric property is represented as a learnable tensor within a fully differentiable optimization framework. The formulation naturally accommodates extensions to multi-frequency inversion and phaseless data reconstruction without requiring structural modifications. Three representative neural operators, FNO, U-FNO, and F-FNO, have been systematically evaluated in terms of reconstruction accuracy and computational efficiency within the proposed framework. The assessment encompasses varying noise levels, single-frequency and multi-frequency data samples, and both with-phase and phaseless measurements. Numerical results demonstrate the generality, robustness, and effectiveness of the proposed framework, confirming its potential as an efficient and versatile solution for complex electromagnetic inverse scattering problems.

\section*{Appendix}

The normalized root mean square error (RMSE)
and the structural similarity index measure (SSIM) \cite{SSIM} are respectively described by

\begin{equation}
\mathrm{RMSE}
= \sqrt{
\frac{\sum \left( \hat{\epsilon}_r - \epsilon_r \right)^2}
     {\sum \left(\epsilon_r \right)^2}
},
\end{equation}

\begin{equation}
\mathrm{SSIM}
= \frac{
\left( 2\mu_{\hat{\epsilon}_r}\mu_{\epsilon_r} + C_1 \right)
\left( 2\sigma_{\hat{\epsilon}_r,\epsilon_r} + C_2 \right)
}{
\left( \mu_{\hat{\epsilon}_r}^2 + \mu_{\epsilon_r}^2 + C_1 \right)
\left( \sigma_{\hat{\epsilon}_r}^2 + \sigma_{\epsilon_r}^2 + C_2 \right)
},
\end{equation}
where $\hat{\epsilon}_r$ denotes the reconstructed relative permittivity,
and $\epsilon_r$ represents the corresponding ground truth distribution. $\mu_{\hat{\epsilon}_r}$ and $\mu_{\epsilon_r}$ are the mean values of the reconstructed
and ground truth permittivity, respectively.
The quantities $\sigma_{\hat{\epsilon}_r}$ and $\sigma_{\epsilon_r}$ denote their standard deviations,
while $\sigma_{\hat{\epsilon}_r,\epsilon_r}$ represents the cross-covariance between them.
The constants are given by $C_1 = (k_1 I)^2$ and $C_2 = (k_2 I)^2$, where $I$ denotes the dynamic range of the permittivity values. The default parameters are $k_1 = 0.01$ and $k_2 = 0.03$.

\ifCLASSOPTIONcaptionsoff
  \newpage
\fi




\bibliographystyle{IEEEtran}
\bibliography{Bibliography}
\vfill

\end{document}